\documentclass[aps,prl,reprint, superscriptaddress,amssymb, amsmath, longbibliography]{revtex4-2}
\usepackage[utf8]{inputenc}
\usepackage{natbib}
\usepackage{comment}
\usepackage{graphicx}
\usepackage{color}
\usepackage{xcolor}
\usepackage[version=4]{mhchem}
\usepackage{gensymb}
\usepackage{xspace}
\usepackage{todonotes}

\newcommand{\ket}[1]{| {#1} \rangle}
\newcommand{\bra}[1]{\langle {#1} |}

\newcommand{\expec}[1]{\left\langle {#1} \right\rangle}
\renewcommand{\vec}[1]{\ensuremath{\boldsymbol{#1}}\xspace}

\newcommand{\brac}[1]{\left[ {#1} \right]}
\newcommand{\tuborg}[1]{\left\{ {#1} \right\}}

\usepackage[
colorlinks=true,     
linkcolor=black,     
citecolor=black,     
urlcolor=blue,       
filecolor=black,     
breaklinks=true,     
pdfusetitle           
]{hyperref}

\usepackage[capitalise,nameinlink]{cleveref}

\begin{document}
	
	\title{Intense-Laser Nondipole-Induced Symmetry Breaking in Solids}
	\author{Asger Weeth}
	\affiliation{Niels Bohr Institute, Copenhagen University, Jagtvej 155A, DK-2200 Copenhagen, Denmark}
	\author{Lars Bojer Madsen}
	\affiliation{Department of Physics and Astronomy, Aarhus University, Ny Munkegade 120, DK-8000 Aarhus C, Denmark}
	\date{\today}

	\begin{abstract}
		High-harmonic spectroscopy in solids gives insight into the inner workings of solids, such as reconstructing band structures or probing the topological phase of materials. High-harmonic generation (HHG)  is a highly non-linear phenomena and simulations guide interpretation of experimental results. These simulations often rely on the electric dipole approximation, even though the driving fields enter regimes that challenge its accuracy. Here, we investigate effects of including nondipole terms in the light-matter coupling in simulations of HHG in materials with both topologically trivial and non-trivial phases. We show how the inclusion of nondipole terms breaks dipole selection rules, allowing for new polarizations of the generated light. Specifically we find that helicity, completely absent in the dipole approximation, is induced by the nondipole extension, and that this helicity is dependent on the topological phase of the material.
	\end{abstract}
	
	\maketitle
	
	With the discovery of high-harmonic generation (HHG) in solids driven at near-infrared wavelengths \cite{ghimire_observation_2011}, a new avenue of strong-field physics opened up. Findings in harmonic spectroscopy of solids include, e.g.,  non-integer harmonics \cite{schmid_tunable_2021, lange_noninteger_2024}, the ability to reconstruct band structures optically \cite{vampa_all-optical_2015}, ultrasensitive spectroscopy \cite{neufeld_ultrasensitive_2019}, and HHG has been proposed as a probe of topological phases \cite{silva_topological_2019, chacon_circular_2020}. For relatively recent reviews on the topic see \cite{ghimire_high-harmonic_2019,Goulielmakis2022}.
	
	When driving solids with intense light with wavelengths ranging into the mid-infrared regime, the field-induced motion of charge carriers gain velocities that are nonnegligible compared to the speed of light, leading to a breakdown of the electric dipole (ED) approximation  \cite{reiss_tunnelling_2014} in a regime different from the domain of short wavelengths, where nondipole (ND) effects are normally considered. Accordingly,  the conditions of HHG call into question the applicability of the ED approximation in this context, and indeed extending the description to include ND contributions has recently been of interest \cite{jensen_nondipole_2020, jensen_beyond_2025}. Such an extension unlocks new and interesting phenomena that can be interpreted in terms of an effective field component in the direction of propagation, which oscillates at double the frequency of the ED component, thus allowing for new transitions \cite{jensen_nondipole_2020, jensen_beyond_2025}. Further more, this new component breaks symmetries which are conserved in the ED limit. These capabilities show the importance of considering ND effects in HHG as these contributions can result in drastically different observables and provide new understanding of the origin of certain spectral features and consequences that are the topic of this work.
	
	In parallel with the interest in emerging ND effects, but without considering these, efforts have aimed at elucidating signatures of topological phase in harmonic spectra (HS) motivated by the ultrafast time-resolving perspective  \cite{chacon_circular_2020, silva_topological_2019, kim_circular_2023, neufeld_are_2023}, and it has been debated whether or not such a signature is unique and universal \cite{neufeld_are_2023}. Based on these studies within the ED approximation,  the best we can say is: "Probably Not", to quote from \cite{neufeld_are_2023}. In order to address this situation, we also here combine studies of topological phase signatures in ultrafast harmonic spectroscopy with consideration of ND effects. The ND field introduces a new vector (the propagation direction of the laser field) that affects the dynamics and could be expected to play a role in topological signatures. To clearly isolate ND and topological effects, we  employ models of both topologically trivial and non-trivial materials. By comparison of spectra generated by these models, we strive to answer the question of whether or not helicity is a unique signature of topological materials. We will investigate the effects of a ND field on graphene and hexagonal boron nitride (hBN), two topologically trivial materials that exhibit different spatial symmetries, which we break with the ND field. The symmetry-breaking capabilities of the ND field is the center of our studies, so we also investigate the topological Kane-Mele model \cite{kane_quantum_2005}, as this model exhibits the dynamical symmetries we want to break with the ND field. We discover that ED selection rules are indeed broken leading to new directions of polarization in the light emitted through the HHG process. We also discover that the symmetry-breaking due to the ND driving laser induces helicity in the emitted light, an entirely new feature completely absent for an ED driving laser and a signature of the ND field. We show that the induced helicity depends on the considered system, and thus encode information about material properties, such as band structures and topological phases. These signatures cannot, however, be readily separated, as they appear in both topologically trivial and non-trivial materials.
	
	A strong-field ND extension of the description of the usual ED light-matter interaction is needed to properly describe the HHG process. Such a description was developed for atoms in \cite{jensen_nondipole_2020}, and recently extended to solid-state systems in \cite{jensen_beyond_2025}. In this approach, the light-matter interaction in the length gauge takes the form
	\begin{equation}
		H_\text{ND}(t) = H_0 + |e|\vec{E}_\text{ND}(t) \cdot\vec{r},
	\end{equation}
	where $H_0$ is the field-free Hamiltonian and the ND contribution is described through the effective electric field 
	\begin{equation}
		\label{NDField}
		\vec{E}_\text{ND}(t) = -\partial_t(\vec{A}^{(0)} (t)+ \vec{A}^{(T)}(t)).
	\end{equation}	
	Here \(\vec{A}^{(0)}(t)\) is the ED term of the vector potential \(\vec{A}(\omega t - \omega z / c)\) propagating along  \(\hat{\vec{z}}\) and \(\vec{A}^{(T)}(t)\)  is given as \(-\partial_t\vec{A}^{(T)}(t) = \frac{|e|}{m}\vec{A}^{(0)} (t)\times \vec{B}^{(1)}(t)\),  with \(e\) the elementary charge, \(m\) the electron mass, and  \(\vec{B}^{(1)}(t) \) the leading-order magnetic-field correction. To examine the impact on symmetries of this new effective field, we consider the vector potential to be linearly polarized along \(\hat{\vec{x}}\) with slowly evolving envelope \(A_0(t)\) and amplitude \(A_0 = E_0/\omega\),
	\begin{align} 
		\vec{A}^{(0)}(t) &=  A_0(t) \sin(\omega t) \hat{\vec{x}} \\
		\vec{A}^{(T)}(t) &= -\frac{|e|}{2mc}(\vec{A}^{(0)}(t))^2 \hat{\vec{z}}.
	\end{align}
	From this field, we can draw conclusions about the impact of a ND driving field. We see that the ND contribution adds a direction of polarization to the field along the direction of propagation, and that this component oscillates at twice the fundamental frequency. As discussed in \cite{jensen_beyond_2025}, these ND effects can be interpreted as a radiation pressure, effecting the trajectories of the electrons in the band structure, and allowing for new \(2n\)-photon transitions between bands. Our attention will, however, be focused on the dynamical symmetries discussed in \cite{neufeld_floquet_2019}, and how the ND field breaks dynamical symmetries not broken by an ED field. These broken symmetries result in new selection rules for the emitted harmonics and thus serve as a signature of the ND field. First, the new directionality of the ND field can break reflection and inversion symmetries present in crystal structures, if the field polarization and propagation directions are chosen properly. Second, the ND field breaks time-reversal (TR) symmetry. As discussed below, breaking this symmetry allows the generated light to be elliptically polarized. Thus, the helicity of the emitted light is a unique ND signature in models where TR is otherwise preserved. 
	
	We consider a solid-state system described by the Hamiltonian  
	\begin{align}
		\mathcal{H} = \sum_{\vec{k}\in \text{BZ}}\sum_{n,m} &a^\dagger_{n\vec{k}}\Big\{H_{0,nm}(\vec{k}) \ +  \nonumber \\
		&|e|\vec{E}_\text{ND}(t)\cdot\brac{\delta_{nm}\vec{\nabla}_{\vec{k}} - i\vec{\xi}_{nm}(\vec{k})}\Big\}a_{m\vec{k}}.
	\end{align}
	Here \(a_{n\vec{k}}(a^\dagger_{m\vec{k}})\) is the fermionic annihilation(creation) operator of a Bloch state with band-index \(n (m)\) at crystal momentum \(\vec{k}\) running through the first Brillouin Zone (BZ), \(H_{0,nm}(\vec{k})\) are the matrix elements of the crystal Hamiltonian, $H_0$, and \(\vec{\xi}_{nm}(\vec{k})\) are the Berry connections. We utilize the Wannier gauge for our simulations. In this gauge the matrix elements read \(H_{0,nm}(\vec{k}) = \sum_{\vec{R}} \exp(i\vec{k}\cdot\vec{R})\bra{\vec{0}n}H_0\ket{\vec{R}m}\) and \(\vec{\xi}_{nm}(\vec{k}) = \sum_{\vec{R}} \exp(i\vec{k}\cdot \vec{R}) \bra{\vec{0}n}\hat{\vec{r}}\ket{\vec{R}m}\) \cite{silva_high_2019}. Here \(\ket{\vec{R}n}\) is the \(n\)'th Wannier state in the cell with lattice vector \(\vec{R}\). The sums run over lattice vectors \(\vec{R}\) between the cells connected within the  considered model \cite{molinero_wannier_2023}. From the many-body current
	\begin{align}
		\expec{\vec{\mathcal{J}} }=  \sum_{\vec{k}\in \text{BZ}}\sum_{n,m}\Big\langle  a^\dagger_{n\vec{k}}\Big\{&\vec{\nabla}_{\vec{k}} H_{0,nm}(\vec{k}) \nonumber\\
		-&i \brac{\vec{\xi}_{nm}(\vec{k}), H_0}_{nm}\Big\}a_{m\vec{k}}\Big\rangle,
	\end{align}
	the harmonic spectrum is obtained via the Fourier relation \(\mathcal{S}(\omega) \propto \omega^2|\mathcal{F}\tuborg{\expec{\vec{\mathcal{J}}}(t)}|^2\). The current is related to the generated field via Lamors formula \cite{Jackson}, neglecting propagation time. We utilize Stokes parameters \cite{Born1999} to determine the helicity of the emitted light. To evaluate \(\expec{\vec{\mathcal{J}}}\),  we need the expectation value of \(a^\dagger_{n\vec{k}}a_{m\vec{k}}\). To this end we consider the reduced density matrix \(\rho_{nm}(\vec{k}, t) = \langle a^\dagger_{m\vec{k}}a_{n\vec{k}}\rangle \) (note the switched indices), whose time-evolution is given by the semiconductor Bolch equations (SBE)
	\begin{align}
		i\hbar\partial_t\rho_{nm}(\vec{k}, t) &= \brac{H_0(\vec{k}), \rho(\vec{k}, t)}_{nm}  \nonumber\\
		& + |e|\vec{E}_\text{ND}(t)\cdot\vec{\nabla}_{\vec{k}}\rho_{nm}(\vec{k}, t)  \nonumber \\
		&- i|e|\vec{E}_\text{ND}(t)\cdot \brac{\vec{\xi}_{nm}(\vec{k}, t), \rho(\vec{k}, t)}_{nm} \nonumber\\
		& - i\hbar \frac{1 - \delta_{nm}}{\tau}\rho_{nm}(\vec{k}, t).
	\end{align}
	The last term accounts phenomenologically for de-phasing time  \(\tau\), usually of a few femtoseconds \cite{vampa_theoretical_2014}. The SBE's were recently cast as the semiconductor Wannier equations \cite{molinero2025}, which offer numerical advantages as well as intuitive interpretations of the electron dynamics in real space. While we will be using the Wannier gauge for our simulations, this implementation of the SBE is also valid for the Hamiltonian gauge \cite{silva_high_2019}. Other methods of circumventing topological obstructions are available, such as utilizing multiple locally smooth gauges \cite{chacon_circular_2020}, or gauge invariant formulations of the SBE \cite{parks_gauge_2023}.

	To investigate the symmetry breaking nature of the ND driving laser, we first address the cases of graphene and hBN. These models consists of electron sites arranged in 2D hexagonal lattices, with nearest-neighbor hopping. Alternating sites \(A\) and \(B\) can be associated with a staggering potential, \(\Delta_i\), \(i = A, B\). For all models we consider, the staggering potentials are chosen as \(\Delta_A = -\Delta_B\). From here on we thus drop the subscript \(B\). The hopping amplitudes are \(t_1f(\vec{k})\), where \(f(\vec{k}) = \sum_{\vec{R}} \exp(-i\vec{k} \cdot \vec{R})\) is a geometric factor. It is important to note, that the vectors \(\vec{R}\) in the Wannier gauge refer to the lattice vectors between unit cells of neighbors, not the vector to said neighbor. The hopping amplitude is \(t_1 = 2.94\) eV in both graphene and hBN \cite{silva_high_2019}. The Hamiltonian of graphene or hBN and the staggering potentials can be found in \cite{Dimitrovski2017}. 
	
	We drive the system with the ND field of \cref{NDField}, with a \(\sin^2\) envelope of 20 cycles. The field is oriented at a grazing angle to the material surface, as to avoid driving the system out of its defined plane.  Our simulations are performed on a \(300\times 300\) Monkhorst-Pack grid \cite{monkhorst_special_1976} of \(k\)-points, initialised with all electrons in the ground state, \(\rho_{11} = 1\), and else \(\rho_{nm} = 0\).  The decoherence time is chosen to be \(\tau = 2\) fs \cite{silva_high_2019}. The field parameters for all simulations are \(E_0 = 4\text{GV}/\text{m}\) and \(\lambda = 3.2\mu \text{m}\), unless otherwise stated and propagating along the \(\Gamma - K\) direction of the lattice. We solve the SBE with a 4th order Runge-Kutta propagator and a time-step of \(2.5\) as. This choice of parameters was verified for convergence.
	\begin{figure}[t]
		\centering
		\includegraphics[width=1\linewidth]{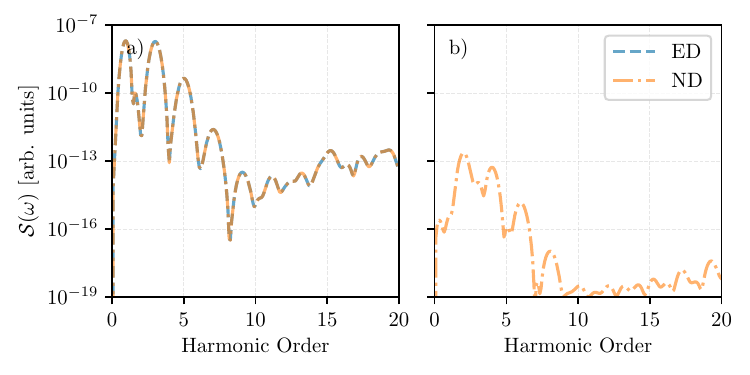}
		\caption{Harmonic spectra of graphene for a driving laser with \(\lambda = 3 \ \mu\)m, polarized along the \(\Gamma-K\) direction. (a)  Spectrum polarized along \(\Gamma - M\) with overlapping ED and ND contributions. (b) Spectrum along \(\Gamma - K\) showing that only ND contributes. The ND driving laser breaks the reflection axis about \(\Gamma - M\), and thus allows for polarization along \(\Gamma-K\).}
		\label{fig:graphenespectra}
	\end{figure} 
	\cref{fig:graphenespectra} shows spectra of graphene driven by ED and ND fields, and the polarization along the high symmetry lines \(\Gamma - M\) and \(\Gamma - K\) are shown. For the ED field, none of the emitted light is polarized along \(\Gamma-K\) due to the reflection symmetry of graphene. The ND driving field breaks this symmetry, allowing for the emitted light to be polarized along \(\Gamma - K\). This ND field also breaks the inversion symmetry of graphene, allowing for even-ordered harmonics, seen in the spectrum along \(\Gamma - K\). These orders appear in the ND spectra due to the ND component oscillating at \(2\omega\), allowing for \(2n\)-photon transitions in the band structure. As we apply a finite pulse of the driving field, the dynamics is not wholly in the Floquet limit, and these even orders are dominant, but not the only contributions. 
	
	\begin{figure}[t!]
		\centering
		\includegraphics[width=1\linewidth]{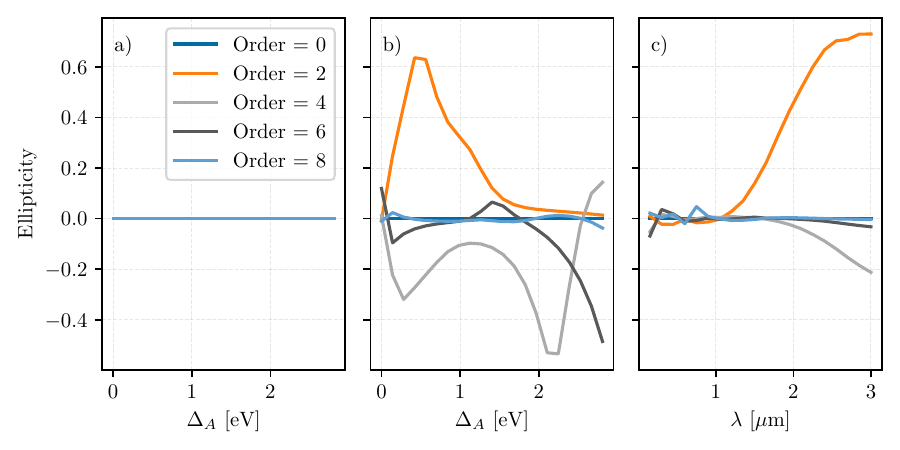}
		\caption{(a) and (b) Ellipticity dependence of the generated light with varying  on-site potential, starting from graphene at \(\Delta_A = 0\) and ending at hBN at \(\Delta_A = 2.81\) eV for ED and ND fields, respectively.  (c) ND ellipticity increase with  driving field wavelength for \(\Delta_A = 0.562\) eV.}
		\label{fig:helicitygraphene}
	\end{figure}
	While these features of inversion symmetry breaking are known from other models \cite{jensen_propagation_2022, jensen_beyond_2025}, we also find a novel feature of the spectra stemming from a ND driving field. Since graphene is TR invariant, we expect only linearly polarized harmonics \cite{neufeld_floquet_2019}. We do, however, break this TR invariance with the ND field, and thus allow for the emitted light to be elliptically polarized. These findings are illustrated in \cref{fig:helicitygraphene}. To investigate the dependence of helicity on material parameters, we vary the on-site potential \(\Delta_A\), scanning from graphene at \(\Delta_A = 0\) eV to hBN at \(\Delta_A = 2.81\) eV and find a striking difference between ED and ND driving fields. We find that the symmetry-breaking nature of the ND driving laser induces helicity in the emitted light, a feature that is forbidden for the ED field. The induced helicity is most pronounced in the even orders, showing that the effect is stemming from the ND field. This helicity is not negligible, reaching values of 0.6 at certain values of the on-site potentials. The helicity depends on this model parameter, with the helicity of multiple orders changing sign as it is varied. Another clear sign of the helicity as a uniquely ND-induced effect can be seen in \cref{fig:helicitygraphene}(c): as the ND component of the effective field compares to the ED component as \(A_\text{ND}/A_\text{ED} = A_0\), the contribution from the former increases with increasing wavelength of the driving field, and so does the induced helicity of the emitted light, showing clearly that this is a ND effect. Thus, the ND driving field leaves a unique and pronounced signature in the spectrum in the form of elliptically polarized harmonics, especially at even harmonics.  As this effect is entirely absent for the ED driving laser, such a signal is a clear indicator of a non-negligible ND contribution.

	\begin{figure}[t]
		\centering
		\includegraphics[width=1\linewidth]{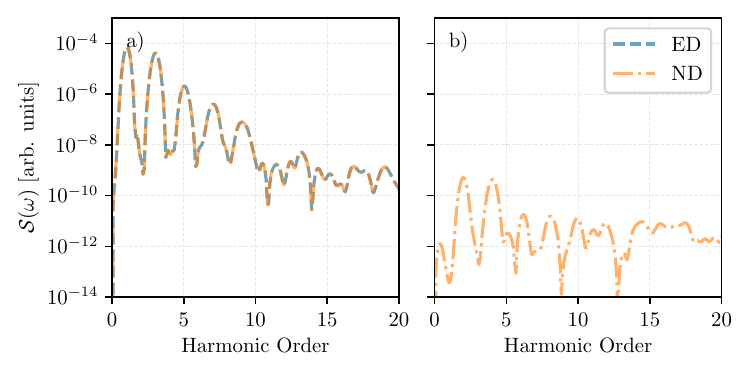}
		\caption{Harmonic spectra of the Kane-Mele model. (a) Spectrum polarized along \(\Gamma-M\) with overlapping ED and ND spectra. (b) Spectrum polarized along \(\Gamma-K\) with only a ND contribution. Like with graphene, the symmetry-breaking nature of the ND field can be seen by breaking the ED selection rule and allowing for the emitted light to be polarized along \(\Gamma-K\).}
		\label{fig:kanemelespectra}
	\end{figure}
	Having discussed graphene and hBN and the clear signature of a ND field, we now discuss whether or not this signature can be used to probe the topological phase, as investigated in detail in \cite{neufeld_are_2023}. We aim to break TR solely with the ND field, so the chosen model system must be TR invariant. This demand excludes Chern insulators described by the Haldane model \cite{haldane_model_1988}: The scalar Haldane model with complex next-nearest-neighbor hopping amplitude intrinsically breaks TR \cite{vanderbilt_berry_2018}. The complex hopping amplitude also defines a direction of hopping, which breaks the spatial symmetries present in graphene and hBN. Thus, there are no clear ED selection rules to be broken by the effective ND field that still, of course, contributes with the radiation pressure and \(2n\)-photon transitions discussed above. We thus seek a model which exhibits spatial and TR symmetries, that can be broken by the ND effective field.  To this end we focus on the Kane-Mele model \cite{kane_quantum_2005}, a quantum spin Hall insulator with \(\mathbb{Z}_2\) topology. This model treats the electrons as proper spinors, each spin being associated with a Chern number, and it is TR invariant~\cite{kane_quantum_2005, soluyanov_wannier_2011}.  The model, like the Haldane model, describes a hexagonal lattice with nearest and next-nearest-neighbor hoppings, and incorporates spin-orbit coupling; see \cite{vanderbilt_berry_2018, qian_role_2022}. We parametrize the model with no spin-spin couplings and control the topological phase via the on-site potentials and the next-nearest neighbour hopping amplitude \(t_2\). The model is in a topologically trivial phase if \(|\Delta_A/(3\sqrt{3}t_2)| > 1\),  and in a \(\mathbb{Z}_2\)-odd topological phase when \(|\Delta_A/(3\sqrt{3}t_2)| \leq 1\). We use the model parameters found in \cite{qian_role_2022}.
	
	\begin{figure}[t]
		\centering
		\includegraphics[width=1\linewidth]{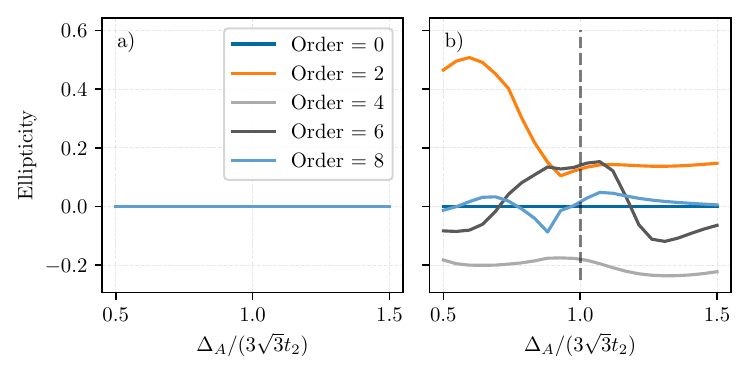}
		\caption{Helicity measurements of the Kane-Mele spectra for ED and ND driving lasers. As the Kane-Mele model is TR invariant like graphene, we break this symmetry with the ND driving laser. In the case of the ED driving laser the helicity is always strictly 0, but for the ND driver it is parameter dependent. The dotted line in (b) indicates the point of topological phase transition.}
		\label{fig:helicitykanemele}
	\end{figure}
	
	The Kane-Mele model shares the reflection symmetry of graphene and hBN, and so the emitted light is only polarized along the \(\Gamma-M\) direction in the case of an ED driving laser propagating along \(\Gamma-K\). \cref{fig:kanemelespectra} shows that this ED selection rule is broken by the ND field. As in graphene, the \(2n\)-photon transitions are evident in the pronounced even harmonics. The breaking of spatial symmetries also allows for harmonics along a new axis. We are, however, mainly interested in the effect of topology on the discovered ND-induced helicity. These results are presented in \cref{fig:helicitykanemele}. As in graphene and hBN, we induce helicity in the emitted harmonics when driving with a ND field. Again, a strong dependence on model parameters can be seen, but no clear signature of topology makes itself apparent. No clear distinction between the two phases, which change at \(\Delta_A/(3\sqrt{3}t_{2}) = 1\), can be seen and thus the topological phase is encoded in the helicity in a non-obvious manner. This is not to say that the topological phase plays no role in the helicity of the emitted light. At the high symmetry points \(K\) and \(K'\), the Berry curvature changes sign with the topological phase, and this sign change together with the new directionality imparted on the electron by the effective ND field effects the electron trajectory in reciprocal space. When varying the model parameters, however, we also make more drastic changes to, e.g., the band structure. As mentioned in \cite{neufeld_are_2023}, these changes have a large impact on the spectrum compared to topological effects, and it is therefore difficult to say which contributions stem from topology and which stem from other sources.
	
	In summary, we have seen how a ND driving field has the ability to break spatial and temporal symmetries, most notably TR symmetry. By breaking these symmetries we are able to break selection rules for harmonic generation, providing a new signature for ND interactions. These new features include both emission of harmonics along ED-forbidden polarizations and the appearance of helicity in the emitted light that is otherwise forbidden by ED selection rules. Through our investigations of graphene, hBN, and the Kane-Mele model, we have seen how especially the helicity of the emitted harmonics reveals the ND extension of the driving field. Furthermore, this helicity measurement is encoded with information about the model in question, as can be seen in the dependence on model parameters, such as the on-site potential. Thus information related to the topological phase of the model will also be encoded in this signal. Other features of the model, like the band gap, are however also dependent on the same model parameters which govern the topological phase. These features will also be encoded in the helicity, and thus this measurement cannot be a unique signature of topology alone. 
	
	\begin{acknowledgments}
		The authors thank Simon Vendelbo Bylling Jensen for fruitful discussions.
	\end{acknowledgments}
	
	\bibliography{ArticleBibliography}
	
\end{document}